# COHERENT RANGING WITH ENVISAT RADAR ALTIMETER: A NEW PERSPECTIVE IN ANALYZING ALTIMETER DATA USING DOPPLER PROCESSING

R. Abileah [a], J. Gómez-Enri [b], A. Scozzari [c], S. Vignudelli [d] *

[a] *jOmegak, San Carlos CA, USA, abileah@jOmegak.com*
[b] *University of Cadíz, Spain, jesus.gomez@uca.es*
[c] *Consiglio Nazionale delle Ricerche (CNR-ISTI), Italy, a.scozzari@isti.cnr.it*
[d] *Consiglio Nazionale delle Ricerche (CNR-IBF), Italy, vignudelli@pi.ibf.cnr.it*

**Abstract:** ESA's Envisat mission carried a RA-2 radar altimeter since its launch in 2002 to sense sea state and especially measure sea surface height (SSH). The onboard processing combined multiple echoes incoherently to reduce Speckle noise and benefit from data compression. In fact, according to past literature the amplitudes were generally expected independent. Nevertheless, samples of complex data time series of individual echoes (IE) were down-linked and archived since 2004 for research studies. In this note we demonstrate that there is sufficient inter-pulse coherence for Doppler processing and we suggest that the archived data can be re-processed into improved SSH. This is of particular interest in challenging domains (e.g., coastal zone) where coherent processing can mitigate errors from ocean surface backscatter inhomogeneity and nearby land backscatter. A new method called zero-Doppler to process IEs is thus proposed and discussed.

* **Corresponding author at:** Consiglio Nazionale delle Ricerche (CNR-IBF) c/o Area della Ricerca CNR, Via Moruzzi 1, 56127 Pisa, Italy – Tel. +39-050-3152804
*E-mail address*: vignudelli@pi.ibf.cnr.it (Stefano Vignudelli)

# 1. INTRODUCTION

The Envisat mission officially ended on 8 April 2012 after almost ten years of successful operation which was twice the planned lifetime. Its legacy is an archive of data from several sensors, one of which is a nadir-pointing radar altimeter (RA-2). The basic concept of the radar altimeter is to transmit a sequence of short pulses of microwave radiation towards the target surface. When operating over the ocean, each single pulse interacts with the rough sea surface and part of the incident radiation reflects back to the radar altimeter. The round trip time to the reflecting surface yields the precise satellite to sea surface distance (known as range). The adopted processing is to range-gate the reflected power by the receiver on board the satellite, that is, to filter the signal into range bins. The series of range-gated amplitudes is usually known as "waveform" in the satellite altimetry community. The time taken to make the round trip between the satellite and the sea surface is deduced from the waveform. Over an ocean surface, the waveform has a characteristic shape that can be described analytically with the Brown model (Brown 1977). The Brown waveform applies to the open-ocean, while there is no comparable model for coastal and inland waters yet. In fact, two problems emerge in coastal waters that are not analytically tractable: inhomogeneous water surface backscatter and land interference. These are well identified and described in Vignudelli et al. (2011) which also reviews the historical development of research in coastal altimetry.

The instantaneous random distribution of the ocean wave facets makes individual real waveforms contain Speckle noise. The noisiness of each individual waveform can be reduced by averaging successive waveforms. Generally, a number of waveforms is added incoherently on board the satellite to give a mean waveform that is then transmitted to the ground, where the subsequent processing steps are performed. A procedure called "retracking" estimates the epoch at mid-height (converted to a range by combination with the computed tracker range), the backscatter (converted to wind speed at sea surface) and the leading edge slope (converted to wave height). Other parameters can be also estimated, including the trailing edge slope (linked to any mispointing of the radar antenna). The retracking procedure is done by fitting the theoretical Brown model to the averaged waveform. The final range measurement is then corrected for atmospheric refraction of the signal propagation in the troposphere and ionosphere as well as the biases generated during the interaction between the electromagnetic radiation and the sea surface. The final product is the sea surface height (SSH) referenced to a geocentric frame. Major details about the satellite altimetry system and the extraction of geophysical parameters from the radar returns can be found in Fu and Cazenave (2001).

The Envisat RA-2 records the backscattered echoes in 128 range-gated bins per pulse. The range gates straddle the expected sea surface range. The early gates (typically 1-45) allow for thermal noise estimation, since the pulse had not yet come back from the water surface. Reflections from interfering scatterers such as ship decks, ocean platforms, and land may also appear in such early gates. The onboard tracker keeps the reflected signal inside the time window of the recorded waveform, trying to tie the leading edge of the echo coming from the air-water interface to a specific location in the time domain, that in the case of Envisat RA-2 corresponds to the bin 46.5 (nominal tracking point). Thus, the leading edge of the typical Brown-like echo rises around bin 46, up to a maximum value, followed by a gradually sloping trailing edge. The radar pulse repetition frequency (PRF) of 1795.332 Hz (generally approximated with 1800 Hz) was specifically chosen to satisfy the Walsh criteria (Walsh 1982), in order to get independent samples. This PRF ensures that a maximum number of independent pulses are available for noise averaging prior to the Brown waveform fitting procedure. Since the pulses are presumed to be independent, the waveforms (formed by 128 range bins) are summed incoherently. Typically 100 raw waveforms are summed, producing averaged waveforms at 18 Hz, which result in measurements at 369 m intervals along the satellite's nadir ground path. In addition, open-ocean applications make frequent use of 1 Hz data, that are obtained by averaging blocks of twenty 18 Hz data produced by the retracking procedure previously mentioned (Gómez-Enri et al. 2009). Such 1 Hz intervals correspond to a distance of about 7 Km between successive along track measurements.

Envisat was one of the last of a series of radar altimeters to use this incoherent power summing method. As first suggested by Raney (1998) radar altimeters can use a mode similar to synthetic aperture radar (SAR) to use the backscatter energy more efficiently, with a significant increase of SSH accuracy. Doppler (or SAR) processing also reduces the effective radar footprint, which is useful near coastlines for the aforementioned reasons. Raney's delay-Doppler technique is the basis of the newest radar altimeter systems: CryoSat-2 (launched on 2010), Sentinel-3 (scheduled to be launched on 2014), Jason-CS (planned for 2017) and SWOT (planned for 2019). A common characteristics of these missions is a much higher PRF of their radar altimeters compared to Envisat, to assure sufficient pulse-to-pulse coherence for Doppler processing. As an example, the CryoSat-2 radar altimeter PRF is 17800, which is nine times higher than Envisat RA-2.

Although the Envisat radar altimeter was designed for incoherent processing, provisions were made for down-linking individual echoes (IE) in bursts of a maximum of 2000 complex echo samples (1.114 second) every 3 minutes (Roca 2005). This recording mode was intended to investigate alternatives to the standard on-board burst summing. Starting from the individual echoes allows other types of processing methods and analyses for open-ocean, coastal, and inland waters. One application of IE data for open-ocean was first suggested by Clifford and Barrick (1978). They derived a theoretical relationship between the ocean significant wave height and a radar altimeter's phase fluctuation spectrum. A radar altimeter, even with uncorrelated pulses, can thus be used for mapping ocean wave heights and Envisat IE data is ideal for testing this use (Gommenginger et al. 2006). In yet another application Berry et al. (2007) suggested that the IE can resolve specular returns from inland water and make possible water level measurements with high spatial resolution, with few or no pulses averaged. However, no prior publications made explicit use of inter-pulse coherency.

In this note we demonstrate that in the Envisat IEs there is pulse-to-pulse coherence for Doppler processing. A new way of processing this data is proposed. This is of particular interest in challenging domains (e.g., coastal zone) where coherent processing can mitigate errors from ocean surface backscatter inhomogeneity and nearby land backscatter. It promises that the archived data can be re-processed into improved geophysical parameters from the radar returns.

## 2. DATA AND METHODS

Envisat IE data packets have been continuously archived since 2004 and are freely available from the European Space Agency in Level 1B products (Berry et al. 2007). A large collection of IE data packets has been cataloged and stored on disk. The orbit is sun-synchronous with a 35-day repeat cycle, thus a given location is revisited every 35 days. The repeatability of the IEs recording every 3 minutes and the orbital stability of the satellite make the burst location in the IE data set quite repeatable on a geographical basis, allowing to create time series of IE-derived geophysical measurements.

One L1B file is a collection of all IE bursts recorded in one satellite orbit (or pass) of cycle 63. One IE record consists of a 2D array of complex (I,Q) values, notated as $z_{n,m}$, where *n=1,...,1984* is the index of consecutive pulses along the satellite track and *m=1,...,128* is the index of concentric range annuli. A further explanation of the concept of range annuli can be found in Fu and Cazenave (2001), where Figure 12 illustrates the sequence of range cells illumination patterns having annular shape.

Therefore, a sum of powers $P_m$ over 100 consecutive pulses, which is similar to the incoherently integrated data (such as those transmitted by the satellite at 18 Hz) can be expressed as follows

$$P_m = \left| \sum_{n \in N} z^*_{n,m} z_{n,m} \right| \qquad (1)$$

where N is a range or subset of 100 consecutive time indices.

For the analysis we use pulse-pair processing (PPP) as described in Miller and Rochwarger (1972). The *k-lag* pulse-pair product, also known as the *k-lag* covariance is

$$S_{k,m} = \sum_{n \in N} z^*_{n,m} z_{n+k,m} \qquad (2)$$

Equation (1) is the *0-lag* pulse-pair product. The phase in the *1-lag* product can be used to measure the Doppler velocity ($V_m$) as follows

$$V_m = \frac{9.91}{\pi} \arg \left| S_{1,m} \right| \qquad (3)$$

with 9.91 being the Doppler Nyquist velocity (in m/s) for the Envisat satellite. The value of 9.91 m/s is the result of multiplying the Nyquist frequency (PRF/2) by the speed of light and dividing by two times the frequency of the Envisat radar altimeter (13.575 $10^9$ Hz). The Doppler velocity should correspond to the satellite's orbital velocity projected on the water surface. If the Doppler spectrum has a Gaussian shape, it has been demonstrated that Equation (3) is not only easier but also more accurate than Fourier analysis for measuring Doppler velocity (Zrnic 1977).

The *k-lag* coherence magnitude is

$$\gamma_{k,m} = \left| S_{k,m} \right| / \left| S_{0,m} \right| \qquad (4)$$

Coherence between echoes makes possible to experiment Doppler processing instead of conventional range altimetry processing. Raney's delayed-Doppler technique (Raney 1998; Jensen 1999) makes the most efficient use of Doppler, by combining both nadir and off-nadir backscatter. The nadir return has a zero Doppler velocity after correction for the satellite's vertical velocity, while the off-nadir returns are brought into focus with an appropriate 'delay-Doppler' correction. This technique is especially suitable for the open ocean. Here we introduce a simplified method that we call zero-Doppler, which is equivalent to Raney's one but using only a zero-delay. The common fundamental requirement for the applicability of such methods is coherence. Compared to delayed-Doppler, zero-Doppler makes a less efficient use of the data because it ignores the off-nadir backscatter. This comes in later range gates that have non-zero Doppler. A reduced land interference is expected, especially if the satellite track is perpendicular to the coastline. In fact, in this case, land is more likely to interfere with higher range bins, being later illuminated by the radar wavefront with respect to the water surface when the nadir point of the satellite is slightly off the coast. Due to the relatively low PRF used in Envisat, the Doppler signals are aliased back on zero-Doppler velocity, however, since the proposed technique ignores off-nadir backscatter these aliases can be easily avoided.

The zero-Doppler velocity technique starts with the same raw complex data, $z_{n,m}$. The phases of the data are adjusted to compensate for the satellite's orbital vertical velocity ($v$), and acceleration ($a$)

$$z'_{n,m} = z_{n,m} e^{-4\pi i f (v t_n + \frac{1}{2} a t_n^2)/c} \tag{5}$$

where $f$ is the radar frequency (13.575 $10^9$ Hz), $c$ is the speed of light, and $t_n$ is the time of the $n^{th}$ pulse. The time is relative to the first pulse in the time series. The ocean surface should now be centered on the zero-Doppler velocity. A low-pass filter is then applied on $z'$ in the time direction to filter out non zero-Doppler velocity signals, producing filtered version *z''*. Experimentation suggests that 8-to-1 bandwidth reduction (1800 Hz filtered down to 225 Hz) provides most of the benefits of Doppler filtering. The waveform is then obtained with coherently summed pulses as follows

$$P_m = \left| \sum_{n \in \mathbb{N}} z''_{n,m} \right|^2 \qquad (6)$$

If there is coherence between echoes the Doppler shift should now align the phases of the complex data so that the echoes sum constructively. If there is no coherence, or the wrong Doppler shift is applied, the phases are random and sum destructively to zero.

## 3. RESULTS

Figure 1 shows a typical example of open-ocean power incoherently integrated waveform (upper panel) and one produced by applying the zero-Doppler velocity method to the same data set (bottom panel). The red and black lines are obtained by averaging 100 IEs (equivalent to RA-2 18 Hz data) and 1800 IEs (thus getting 1 Hz data), respectively. Observation of the two panels highlights the different waveform responses around the common tracking point. As expected, Doppler filtering sharpens the waveform response at the bins that correspond to the surface at nadir. The secondary peaks are aliases. The Doppler velocity increases monotonically at ranges beyond the nadir, where the instantaneous vertical component of the satellite velocity is the only contribution to the Doppler shift. As seen in Equation (3), Doppler velocities greater than the 9.91 m/s Nyquist limit produce aliases. As a consequence, at certain ranges the aliasing folds the Doppler back on zero-Doppler velocity, so that secondary peaks appear.

According to the Brown theory, the leading slope of the waveform described analytically by the Brown model is inversely proportional to the significant wave height (SWH). Under the hypothesis of open-ocean conditions, the leading slope of the collected raw waveforms conforms to what is expected from the Brown model. Therefore, the leading slope of the waveforms analyzed here is assumed as an indicator of the sea state, which is quantified by the SWH. Coherence is expected to be highly correlated with SWH, being least in high sea states, and higher in low sea state, thus smoother water surfaces. The coherence magnitudes were estimated by applying Equation (4) to 1263 open-ocean records from one month covering North and South Pacific. The results show that the coherence is greatest at one-lag but surprisingly is often statistically significant at two-, three- and four-lags (not shown in this paper).

Figure 2 is a plot of the coherence magnitudes computed at one-lag against the leading slope of the collected waveforms for all the 1263 records. The coherence values are based on summing over a 1.114 second record (i.e., n=1,...,1984) that is equivalent to averaging over about 7 km along track. The slope is normalized to an arbitrary scale in the range of 0 to 100. The plot shows the increasing coherence with decreasing SWH (i.e. higher leading edge slope), confirming the expected dependence. Coherence magnitudes are from 0.05 to 0.3, with 0.2 being the most typical value. The expected coherence magnitude is $1/\sqrt{N}$ with a standard deviation of $1/2\sqrt{N}$. A coherence magnitude around 0.2 is thus statistically very significant when N is 1984. Although it appears to contradict the Walsh theorem and the generally held views on the statistical independence of Envisat IEs, the correlation between the coherence magnitude and the leading slope of the collected waveforms highlights that coherence is not an artifact produced by the radar system nor the signal processing chain.

The inter-pulse coherence is further confirmed by verifying that the Doppler velocity should be measuring the satellite's vertical orbit velocity. Figure 3 shows the scatterplot of Doppler velocity estimates in the 1263 data records (using Equation 3) against the known Doppler velocity, which is based on determination from precise orbital mechanics and available in the IE records. This is a further proof that there is coherence between echoes. The correlation is substantially linear with a root mean square dispersion of about 0.2 m/s around the linear relationship. About half the time the vertical orbital velocity is outside the Doppler velocity Nyquist limits ($\pm 9.91$ m/s). In those cases the measured Doppler is aliased into the $\pm 9.91$ m/s interval.

The above findings enable the application of the Doppler processing to the IE records. Coming back to Figure 1 (lower panel), the first peak in the waveform is all that matters for the range assessment. The secondary peaks are due to aliasing of energy that a zero-Doppler velocity filter would have eliminated if the PRF was higher. Although Envisat RA-2 was not designed for Doppler processing, such aliases are easy to identify because after zero-Doppler velocity they are found at zero-Doppler shifts in higher range bins.

A preliminary check about the real capability of this processing approach to mitigate the effect of ocean surface inhomogeneity and nearby land scatterers has also been performed. Figure 4 shows one example to illustrate the zero-Doppler velocity technique in coastal waters. In this case-study, the satellite track is almost perpendicular to the coastline of Make-Jima, a small island in Japan. In the incoherent radar altimeter waveform there is evidence of land contamination with respect to the Brown model, being characterized by a typical double-peaking feature (Figure 5, upper panel). The zero-Doppler velocity method produces a much improved waveform by reducing the land interference (Figure 5, lower panel). The waveforms shown in the two panels have been obtained by averaging N=1984 IEs in a single burst.

## 3. DISCUSSION AND CONCLUDING REMARKS

An eight year archive of Envisat IE data is available for scientific research and its exploitation until now has been limited. Envisat was by design supposed to produce statistically independent echoes. The provision of IE samples was to test pulse to pulse coherence in view of the next generation of altimeters, however, the potential for Doppler processing with those IE data has not been sufficiently appreciated. However, this note demonstrates that the typical inter-pulse coherence magnitude levels are ~0.2, that with 1984 samples is significant for Doppler processing. A zero-Doppler velocity method in the form of a simplified version of Raney's delay-Doppler altimeter is proposed. The results show that the zero-Doppler velocity processing has the capability to filter out land interference. This was anticipated from Raney's analysis but now it is confirmed with IE data produced by Envisat RA-2.

The findings demonstrated in this note open the way to new possibilities in processing and exploiting IE datasets, stimulating the development of further research. The zero-Doppler method is easy to implement and particularly appropriate for open-ocean applications, and also very promising for challenging domains such as the coastal zone. The method is also relevant to inland water bodies, which are often characterized by a limited number of pulses over water. Coherent processing enables more efficient use of those fewer pulses. The smaller the water body the more it makes sense to do zero-Doppler instead of delayed Doppler method because only the near nadir range gates matter.

This study also enables a new view point about the meaning and usefulness of having a flexible time integration. Despite the fixed time period that characterized products associated with conventional satellite altimetry, the integration time has to be adjusted to application and location. In particular, by using IEs produced by Envisat RA-2 a reasonable strategy can be: (1) very long coherent periods over open-ocean up to the scale of the local SSH; (2) longer periods if the track is parallel to the coastline; (3) shorter periods when the track is perpendicular to the coastline in order to cut the portion of the record mostly contaminated by the land and helping with the surface wind inhomogeneity or with resolving near shore SSH slope; (4) variable periods dictated by the width of the inland water body (e.g. rivers) and the duration of the specular return.

Further studies, that fall outside the scope of this note, would be required in order to truly establish the ability of the proposed method to improve coastal or inland water measurements. A further development of this study may regard the collection of a large number of coastal and inland water samples, in order to form adequate statistics, implying a detailed analysis in each case of land mass entering the tracking window of the altimeter.

In addition, further research is needed to demonstrate that IE data provides the promised improvements in water level measurements as anticipated by Jensen and Raney (1998). In fact, the extraction of the range information can use the range bins on the leading edge and on the backside of the first peak in a similar fashion to the proposed approach for SAR signals, thus representing an attractive step further in the development and exploitation of this technique.

CryoSat-2 is already in operation to provide Doppler processing since 2010 at much higher PRF than Envisat, and this represents a distinct advantage. Sentinel-3's launch in 2014 will add further Doppler processing capability and high data rate. However, Envisat's archive, extending back to 2004, will continue to be of great scientific value. Envisat IEs have the unique interleaving potential of comparing conventional and Doppler processing at same place and time. Lesson learned by this precious dataset can provide guidance for coming missions designed with ocean in mind.

# ACKNOWLEDGMENTS

The analysis of Envisat IE data has been inspired by the ESA funded COASTALT project, that was aimed at exploiting Envisat altimeter data in the coastal zone. The help of several individuals is graciously acknowledged. Salvatore Dinardo and Jérôme Benveniste of European Space Agency made possible the access to the IE archive. Roberto Cuccu of European Space Agency helped in downloading data through the GPOD service. We consulted several experts in the field after our initial analysis demonstrated what was initially an unexpected pulse-to-pulse coherence in IE data. Walter Smith of NOAA brought the Walsh criterion to our attention and suggested a crucial test of comparing the measured Doppler velocity with the satellite vertical orbit velocity. Keith Raney of JHU/APL provided further interesting discussion and comments about Doppler processing . The results in this paper were first reported at the 6th Coastal Altimetry Workshop, Riva del Garda, Italy, 20-21 September (www.coastalt.eu). The authors would like also to thank two anonymous reviewers for their valuable suggestions and comments to improve the quality of the paper and stimulate the exploitation of Doppler processing in satellite radar altimetry. This study was partially supported by the Spanish Ministry of Education and Science (CGL2008-04736).

# REFERENCES

Berry P. A. M., Freeman, J. A., Rogers, C., & Benveniste, J. (2007). Global Analysis of Envisat RA-2 Burst Mode Echo Sequences. *IEEE Transactions on Geoscience and Remote Sensing*, Vol. 45, No. 9, 2869-2874, doi:10.1109/TGRS.2007.902280.

Brown, G. S. (1977). The Average Impulse Response of A Rough Surface and Its Applications. *IEEE Transactions on Antennas and Propagation,* Vol. 25, No. 1, 67-74, doi:10.1109/TAP.1977.1141536.

Clifford, S.F., & Barrick, D. E. (1978). Remote sensing of sea state by analysis of backscattered microwave phase fluctuations. *IEEE Transactions on Antennas and Propagation*, Vol. 26, No. 5, 699-705, doi:10.1109/TAP.1978.1141915.

Fu, L. L., & Cazenave, A. (2001). *Satellite altimetry and Earth sciences, A Handbook of techniques and applications*. International Geophysics Series, 69, Academic Press, San Diego, 463 pp.

Gómez-Enri, J., Vignudelli, S., Quartly, G., Gommenginger, C., & Benveniste, J. (2009). Bringing satellite radar altimetry closer to shore. *SPIE (Society of Photo-Optical Instrumentation Engineers) Newsroom*, 1-3, doi: 10.1117/2.1200908.179.

Gommenginger, C., Challenor, P., Gómez-Enri, J., Quartly, G., Srokosz, M., Berry, P.A.M., Garlick, J.D., Cotton, D., Carter, D., Rogers, C, Haynes, S., LeDuc, I., Milagro M. P., & Benveniste, J. (2006). New scientific applications for ocean, land and ice remote sensing with ENVISAT altimeter individual echoes. *Proceedings of the Symposium on 15 Years of Progress in Radar Altimetry 13-18 March 2006*, European Space Agency Publication SP-614, ESTEC, Postbus 299 2200 AG Noordwijk, The Netherlands.

Jensen, J. R & Raney, R.K. (1998). Delay/Doppler Radar Altimeter: Better Measurement Precision. *Proceedings of the IEEE Geoscience and Remote Sensing Society (IGARSS) Symposium 6-10 July 1998*, Vol. 4, 2011-2013, doi: 10.1109/IGARSS.1998.703724.

Jensen, J. R. (1999). Radar Altimeter Gate Tracking: Theory and Extension. *IEEE Transactions on Geoscience and Remote Sensing*, Vol. 37, No. 2, 651-658, doi:10.1109/36.752182.

Miller, K., & Rochwarger, M. (1972). A Covariance Approach to Spectral Moment Estimation. *IEEE Transactions on Information Theory*, Vol. 18, No. 5, 588-596, doi: 10.1109/TIT.1972.1054886.

Raney, R. K. (1998). The Delay/Doppler Radar Altimeter. *IEEE Transactions on Geoscience and Remote Sensing*, Vol. 36, No. 5, 1578-1588, doi:10.1109/36.718861.

Roca, M. (2005). *The EnviSat RA-2 Individual Echoes (or Burst Mode) - Level 0 & Level 1b Processing - Detailed Processing Model (DPM)*. Report Pildo Consulting S.L., 46 pp (Available on: http://earth.esa.int/raies/docs/IE_DPM.Is2a.fm.pdf).

Vignudelli, S., Kostianoy, A.G., Cipollini, P., & Benveniste, J. (2011) *Coastal Altimetry*, Springer-Verlag Berlin Heidelberg 2011, 565 pp, doi:10.1007/ 978-3-642-12796-0.

Zrnic, D. S. (1977). Spectral Moment Estimates from Correlated Pulse Pairs. *IEEE Transactions On Aerospace And Electronic Systems*, Vol. 13, No. 4, 344-354, doi: 10.1109/TAES.1977.308467

Walsh E. J. (1982), Pulse to pulse correlation in satellite radars, *Radio Science*, Vol. 17, No. 4, 786–800, doi:10.1029/RS017i004p00786.

**FIGURES**

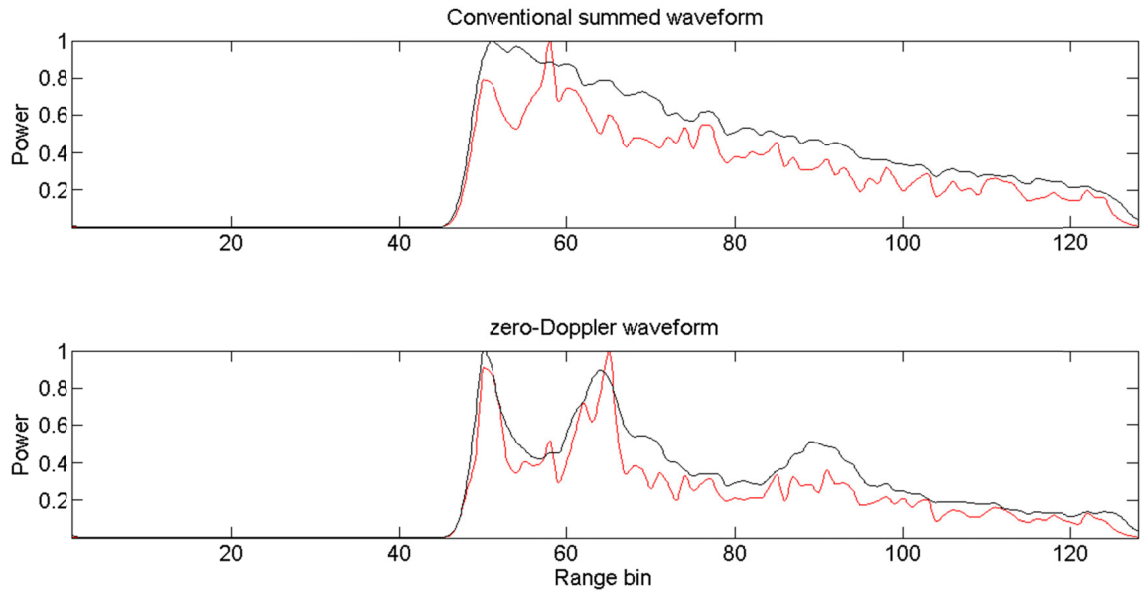

**Figure 1:** Example of Envisat waveforms over open-ocean. Data taken from pass 063, cycle 63. Upper Panel: Incoherently integrated waveforms (Black=1 Hz and Red=18 Hz). Bottom Panel: the corresponding waveforms produced by the Doppler Processing (zero-lag).

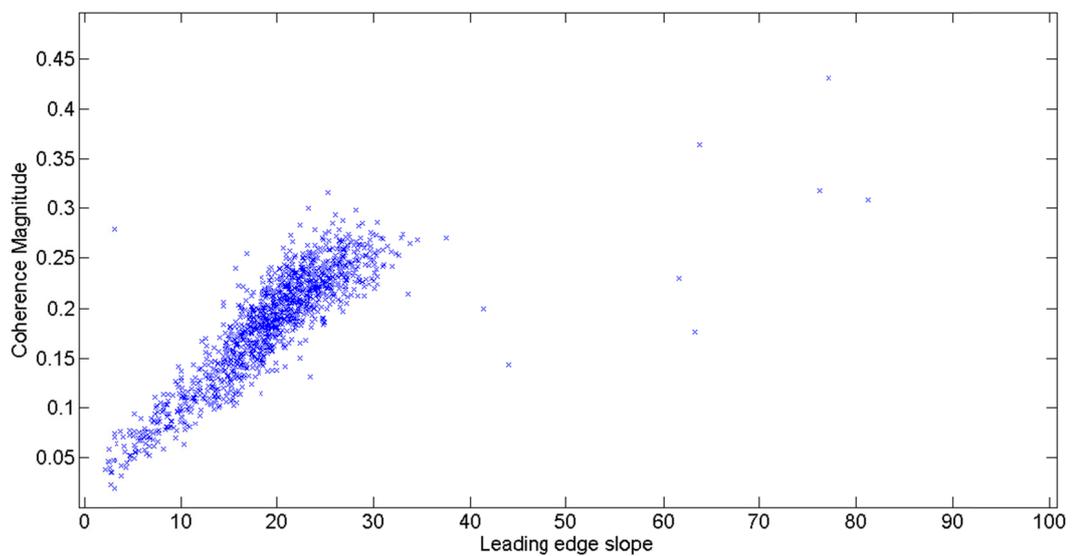

**Figure 2:** Coherence magnitudes against leading slopes for all 1263 over open-ocean records from one month covering North and South Pacific.

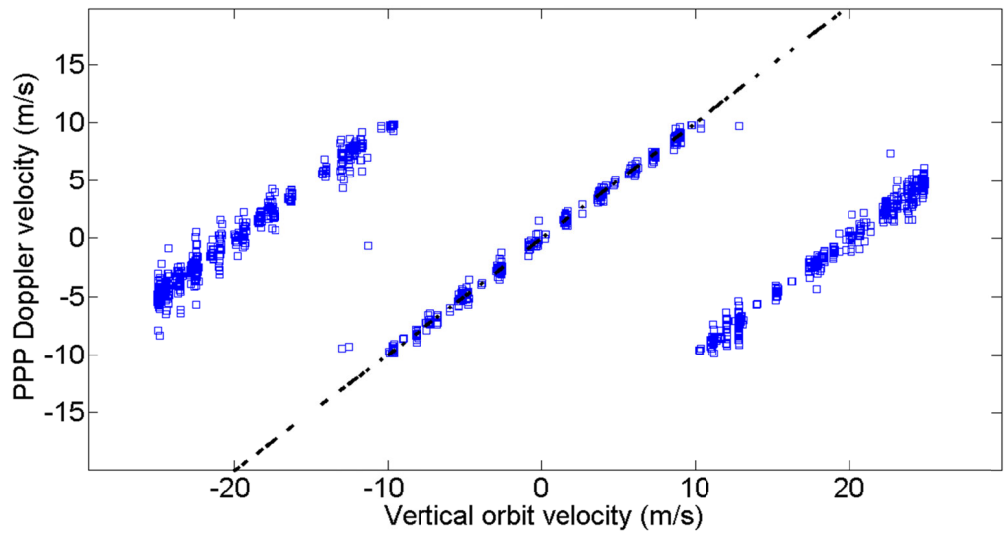

**Figure 3:** Doppler velocity estimated by Equation (3) against the satellite vertical velocity determined from precise orbital mechanics and available in the IE records. Data are from 1263 over open-ocean records from one month covering North and South Pacific. Doppler velocities outside the Nyquist velocities (- 9.91 to + 9.91 m/s) are aliased.

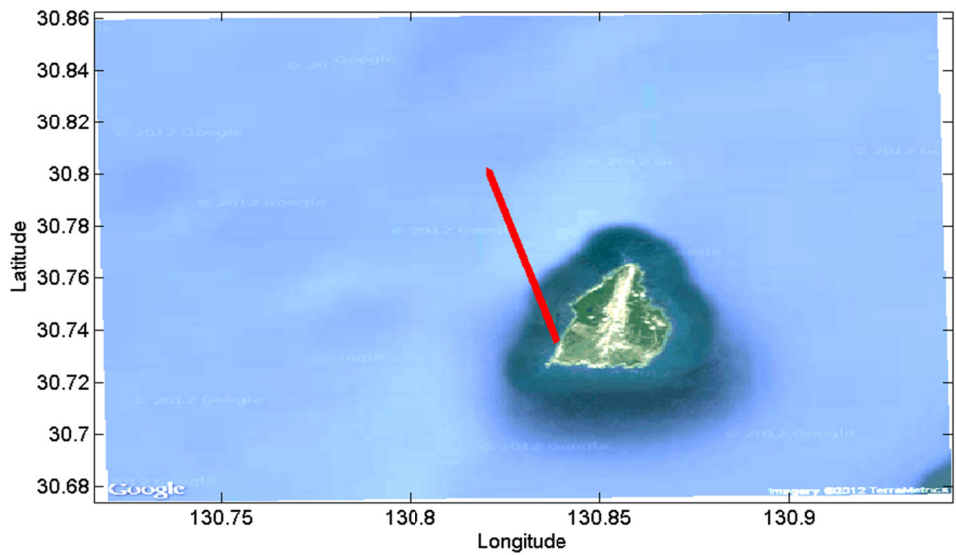

Figure 4: Orbit segment corresponding to one IE burst collected by Envisat near Make-Jima, a small island in Japan.

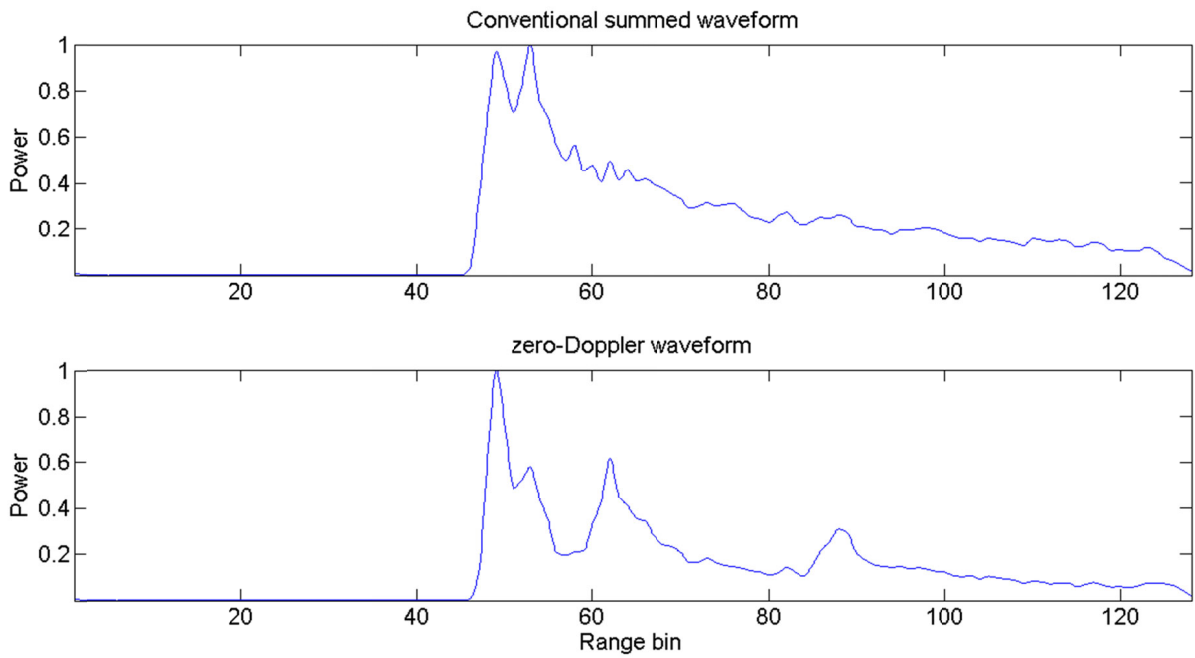

Figure 5: Comparison of conventional (upper panel) and zero-Doppler (lower panel) radar altimetry waveforms for IE data just off the coast of Make-Jima, a small island in Japan (see Figure 4). Data taken from pass 381, cycle 63.